# A new methodology for constructing a publication-level classification system of science

Ludo Waltman and Nees Jan van Eck

Centre for Science and Technology Studies, Leiden University, The Netherlands
{waltmanlr, ecknjpvan}@cwts.leidenuniv.nl

Classifying journals or publications into research areas is an essential element of many bibliometric analyses. Classification usually takes place at the level of journals, where the Web of Science subject categories are the most popular classification system. However, journal-level classification systems have two important limitations: They offer only a limited amount of detail, and they have difficulties with multidisciplinary journals. To avoid these limitations, we introduce a new methodology for constructing classification systems at the level of individual publications. In the proposed methodology, publications are clustered into research areas based on citation relations. The methodology is able to deal with very large numbers of publications. We present an application in which a classification system is produced that includes almost ten million publications. Based on an extensive analysis of this classification system, we discuss the strengths and the limitations of the proposed methodology. Important strengths are the transparency and relative simplicity of the methodology and its fairly modest computing and memory requirements. The main limitation of the methodology is its exclusive reliance on direct citation relations between publications. The accuracy of the methodology can probably be increased by also taking into account other types of relations, for instance based on bibliographic coupling.

## 1. Introduction

In bibliometric and scientometric research, classification systems of science are an indispensable tool. A classification system of science assigns journals or individual publications to research areas. Such a system can for instance be used to simplify literature search, to study the structure and dynamics of scientific disciplines, or to facilitate bibliometric research evaluations.

This paper introduces a new methodology for constructing a classification system of science. The core of the proposed methodology consists of a large-scale clustering of scientific publications. Publications are clustered based on citation relations. Each publication is assigned to a single research area, and research areas are organized in a hierarchical structure. At the highest level, research areas may for instance correspond with broad scientific disciplines. At the lowest level, they may correspond with small subfields. The proposed methodology is able to cluster very large numbers of publications. In the application presented in this paper, a clustering of almost ten million publications is produced. This application shows that the proposed methodology can be used to construct a classification system that includes essentially all publications in the international scientific literature in a time period of several years.

There are many different classification systems of science. For bibliometric and scientometric purposes, the most popular classification system is without doubt the system included in Thomson Reuters' Web of Science database. This system consists of about 250 research areas, referred to as subject categories. A somewhat similar system is included in Elsevier's Scopus database. The classification systems of Web of Science and Scopus work at the level of scientific journals. In these systems, a journal is assigned to one or more research areas. Publications are not directly assigned to research areas. Instead, the journal in which a publication has appeared determines the research area(s) to which the publication belongs.

In addition to the classification systems of Web of Science and Scopus, researchers have been working on developing their own classification systems (Archambault, Beauchesne, & Caruso, 2011; Glänzel & Schubert, 2003; Klavans & Boyack, 2010) or on testing techniques that can be used to construct such systems (e.g., Rafols & Leydesdorff, 2009; Rosvall & Bergstrom, 2008, 2011). Research attention has focused mainly on classification systems that work at the level of journals, just like the systems of Web of Science and Scopus. Journal-level classification systems normally consist of at most a few hundred research areas, and they need to deal with multidisciplinary journals such as *Nature* and *Science* in a special way, for instance by assigning these journals to a special category for multidisciplinary sources. Compared with their journal-level counterparts, publication-level classification systems, which work at the level of individual publications, have received less attention in the literature. Early work on publication-level classification systems, at a relatively small scale, was done by Small and colleagues (Griffith, Small, Stonehill, & Dey, 1974; Small & Griffith, 1974; Small & Sweeney, 1985; Small, Sweeney, & Greenlee, 1985). Recently, a large-scale publication-level classification system was constructed by Klavans and Boyack (2010). In this system, more than 5.5 million publications have been assigned to over 84,000 research areas.

Compared with earlier work on constructing classification systems of science, the methodology that we introduce in this paper has a number of advantages. First of all, our methodology works at the level of individual publications rather than at the journal level. This allows for a more detailed classification of science, and it avoids difficulties with multidisciplinary journals. A second advantage of our methodology is its ability to handle very large numbers of publications. The application that we present in this paper shows how our methodology produces a clustering of almost ten million publications, and if needed even larger numbers of publications could be handled. The number of publications that we deal with exceeds other recently published large-scale clustering analyses (e.g., Boyack & Klavans, 2010; Boyack et al., 2011; Klavans & Boyack, 2010). A third strength of our methodology is its transparency and relative simplicity combined with the minimal amount of human involvement it requires. Our methodology is described in full detail in this paper, and using this description as well as the software that we have made freely available, it should be possible for anyone with sufficient data access and sufficient computing resources to replicate the steps we take. Human involvement is minimized to the choice of suitable values for a small number of parameters.

The organization of this paper is as follows. Section 2 provides a detailed description of our proposed methodology. Section 3 introduces an application in which we use our methodology to construct a classification system that includes almost ten million publications from the period 2001–2010. Section 4 presents an analysis of the resulting classification system. Finally, Section 5 concludes the paper by discussing some limitations of our methodology and some directions for future research.

## 2. Methodology

The methodology that we propose for constructing a publication-level classification system of science can be subdivided into three steps:
1. Determining the relatedness of publications.
2. Clustering publications into research areas.
3. Labeling research areas.

These steps are discussed in detail in the following three subsections.

**2.1. Step 1: Determining the relatedness of publications**

In the first step of our methodology, we start with a set of publications and we determine the relatedness of each pair of publications in this set. Let *n* denote the number of publications, and let $c_{ij} \geq 0$ denote the relatedness of publications *i* and *j*. In principle, the relatedness of publications can be determined in many different ways, for instance based on direct citations, co-citations, bibliographic coupling, shared words in titles and abstracts, or a combination of these elements. However, when working with millions of publications, it is crucial to minimize as much as possible the number of publication pairs for which $c_{ij}$ is greater than 0. This saves computer memory and reduces computing time. For this reason, we use only direct citations from one publication to another to determine the relatedness of publications. We disregard the direction of a citation. In this way, a simple binary definition of $c_{ij}$ is obtained: $c_{ij}$ equals 1 if either publication *i* cites publication *j* or publication *j* cites publication *i*, and $c_{ij}$ equals 0 if there is no direct citation relation between publications *i* and *j*.

**2.2. Step 2: Clustering publications into research areas**

Our methodology produces hierarchical classification systems. Each publication belongs to a single research area at the lowest level of a classification system, each research area at the lowest level in turn belongs to a single research area at the second-lowest level, and so on.

In the second step of our methodology, we use the results from the first step to build the basic structure of a classification system. This means that publications are clustered into research areas and that research areas are organized in a hierarchical structure.

The second step of our methodology involves a small number of parameters. Suitable values for these parameters depend on the purpose for which a classification system is intended to be used. The most basic parameter, denoted by *L*, is the number of levels in a classification system. We refer to the highest level of a classification system as level 1, the second-highest level as level 2, and so on. The lowest level of a classification system is referred to as level *L*. For each level *l* in a classification system, there are two additional parameters: The resolution parameter, denoted by $r^{(l)}$, and the minimum number of publications per research area, denoted by $n_{\min}^{(l)}$. The resolution parameter $r^{(l)}$ determines how much detail is offered at level *l* in a classification system. The higher the value of this parameter, the larger the number of research areas at level *l* and the smaller the average number of publications per research area. The resolution parameters $r^{(1)},\ldots,r^{(L)}$ must satisfy the following condition:

$$0 \leq r^{(1)} < \ldots < r^{(L)} \leq 1. \tag{1}$$

The parameter $n_{\min}^{(l)}$ ensures that each research area at level $l$ includes at least a certain minimum number of publications.

It is well known that there are large differences in citation behavior among scientific fields. Because of these differences, the relatedness scores obtained in the first step of our methodology cannot be directly compared across fields. Publications in one field (e.g., cell biology) may for instance tend to have a much higher total relatedness score than publications in another field (e.g., mathematics). Such differences among fields may lead to an unbalanced classification system in which some fields are overrepresented while others are underrepresented. To correct for differences among fields, relatedness scores need to be normalized. The normalization that we perform is similar to the idea of fractional citation counting first proposed by Small and Sweeney (1985). We define the normalized relatedness of publication $i$ with publication $j$ as

$$a_{ij} = \frac{c_{ij}}{\sum_k c_{ik}}. \tag{2}$$

Notice that this definition does not require normalized relatedness scores to be symmetric, that is, $a_{ij}$ need not be equal to $a_{ji}$. According to (2), the normalized relatedness of publication $i$ with publication $j$ equals the relatedness of the two publications divided by the total relatedness of publication $i$ with all other publications. It follows from (2) that for each publication the total normalized relatedness with all other publications equals one.[1] Hence, in a sense (2) ensures that all publications have the same overall weight. This guarantees that differences among fields are corrected for.

We take a bottom-up approach to build the structure of a classification system. Our approach consists of a number of iterations, one for each level of a classification system. The lowest level of a classification system is constructed in the first iteration, the second-lowest level in the second iteration, and so on. Each iteration involves two stages. In the first stage, a preliminary assignment of publications to research areas is made. After this stage, it may be that some research areas include fewer publications than the minimum required by the parameter $n_{\min}^{(l)}$. This is corrected in the second stage, in which research areas with an insufficient number of publications are discarded and the publications belonging to these areas are reassigned to other areas.

We use $x_i^{(l)}$ to denote the preliminary assignment of publication $i$ to a research area at level $l$ in a classification system. The final assignment of publication $i$ to a research area at level $l$ is denoted by $y_i^{(l)}$. In order to obtain a classification system with a proper hierarchical structure, we require that for all publications $i$ and $j$ and all levels $l \in \{2, \ldots, L\}$ the following condition is satisfied:

$$y_i^{(l)} = y_j^{(l)} \Rightarrow x_i^{(l-1)} = x_j^{(l-1)}. \tag{3}$$

---

[1] There is one exception. If a publication $i$ does not have any relation with other publications, the denominator in (2) equals zero. In this case, we set $a_{ij}$ equal to zero for all publications $j$, which means that the total normalized relatedness of publication $i$ with all other publications equals zero as well.

This condition ensures that if two publications belong to the same research area at level $l$ in a classification system, they also belong to the same research area at all higher levels in the system.

As already mentioned, the structure of a classification system is built in a bottom-up fashion. We start at the lowest level (i.e., level $L$) and then move up one level at a time until the highest level (i.e., level 1) has been reached. In the first stage of each iteration, a clustering technique is employed to obtain a preliminary assignment of publications to research areas. The clustering technique that we use assigns publications to research areas by searching for values of $x_1^{(l)},\ldots,x_n^{(l)}$ that maximize the following quality function:

$$V^{(l)}(x_1^{(l)},\ldots,x_n^{(l)}) = \sum_i \sum_j \delta(x_i^{(l)}, x_j^{(l)})(a_{ij} - r^{(l)}). \qquad (4)$$

In this function, $\delta(x_i^{(l)}, x_j^{(l)})$ equals 1 if $x_i^{(l)} = x_j^{(l)}$ and 0 otherwise. At each level $l$ except the lowest, maximization of (4) is performed subject to the constraint in (3). The quality function in (4) ensures that a publication is assigned to a research area only if it is sufficiently related to the other publications belonging to that area. Whether publications are considered to be sufficiently related depends on the resolution parameter $r^{(l)}$. The higher the value of $r^{(l)}$, the more strongly publications need to be related in order to be assigned to the same research area. Because of this mechanism, a higher value of $r^{(l)}$ tends to lead to a larger number of research areas and, consequently, to a more detailed classification system.

Clustering techniques similar to ours have been studied extensively in the network analysis literature. The most popular technique is modularity-based clustering (Newman, 2004a, 2004b; Newman & Girvan, 2004). Many different variants of modularity-based clustering have been proposed in the literature. The idea of a resolution parameter was introduced in a variant proposed by Reichardt and Bornholdt (2006). Recently, a variant of modularity-based clustering that uses exactly the same quality function as in (4) was proposed by Traag, Van Dooren, and Nesterov (2011). The clustering approach that we take is also closely related to our own earlier work (Waltman, Van Eck, & Noyons, 2010). The only difference is that in our earlier work we did not use the same normalization as in (2).

Our clustering technique requires an optimization algorithm to search for values of $x_1^{(l)},\ldots,x_n^{(l)}$ that maximize the quality function in (4). There is a considerable literature on optimization algorithms for modularity-based clustering and its variants. One class of algorithms are the so-called multilevel local search algorithms. An elaborate analysis of the performance of different algorithms belonging to this class is presented by Rotta and Noack (2011). The optimization algorithm that we use is inspired by the work of Rotta and Noack, but it also includes some further extensions. In the terminology of Rotta and Noack, our algorithm is based on a combination of multilevel coarsening and multilevel refinement. Multilevel coarsening is also the basic mechanism employed in the well-known algorithm of Blondel, Guillaume, Lambiotte, and Lefebvre (2008),[2] but this algorithm does not include a multilevel refinement mechanism. We have implemented our optimization algorithm in a freely available computer program (see www.ludowaltman.nl/classification_system/). The C source code of the algorithm is available as well.[3]

---

[2] For examples of bibliometric/scientometric studies in which this algorithm is used, we refer to Colliander and Ahlgren (2012), Rafols and Leydesdorff (2009), Wallace, Gingras, and Duhon (2009), and Zhang, Liu, Janssens, Liang, and Glänzel (2010).

We emphasize that finding values of $x_1^{(l)},\ldots,x_n^{(l)}$ for which the quality function in (4) is maximized is a difficult task, especially when working with large numbers of publications. Exact maximization of (4) is usually not feasible. Instead, our optimization algorithm aims to find values of $x_1^{(l)},\ldots,x_n^{(l)}$ for which (4) is at least close to its maximum. Because our algorithm includes some random elements, different runs of the algorithm will generally lead to different values of $x_1^{(l)},\ldots,x_n^{(l)}$. To get as close as possible to the maximum of (4), the algorithm can be run multiple times. The values of $x_1^{(l)},\ldots,x_n^{(l)}$ resulting from the run in which the highest value of (4) is obtained can then be kept, while the values of $x_1^{(l)},\ldots,x_n^{(l)}$ resulting from the other runs of the algorithm can be discarded. In this way, the larger the number of runs of the algorithm, the closer one will get to the maximum of (4).

The preliminary assignment of publications to research areas provided by $x_1^{(l)},\ldots,x_n^{(l)}$ may lead to research areas that include fewer publications than the minimum required by the parameter $n_{\min}^{(l)}$. This is why we need a second stage in each iteration of our bottom-up approach. In this second stage, research areas with an insufficient number of publications are discarded and the publications belonging to these areas are reassigned to other areas.

Let $\bar{a}_{uv}^{(l)}$ denote the relatedness of research areas $u$ and $v$ based on the preliminary assignment of publications to research areas provided by $x_1^{(l)},\ldots,x_n^{(l)}$. We define the relatedness of two research areas as the average normalized relatedness of the publications belonging to the two areas. Hence, $\bar{a}_{uv}^{(l)}$ is given by

$$\bar{a}_{uv}^{(l)} = \frac{\sum_i \sum_j \delta(x_i^{(l)},u)\delta(x_j^{(l)},v)a_{ij}}{\sum_i \sum_j \delta(x_i^{(l)},u)\delta(x_j^{(l)},v)}, \qquad (5)$$

where $\delta(x_i^{(l)},u)$ equals 1 if $x_i^{(l)} = u$ and 0 otherwise. Let $S^{(l)}$ denote the set of all research areas that include at least $n_{\min}^{(l)}$ publications. More formally,

$$u \in S^{(l)} \Leftrightarrow \sum_i \delta(x_i^{(l)},u) \geq n_{\min}^{(l)}. \qquad (6)$$

Publications with a preliminary assignment to a research area in the set $S^{(l)}$ do not need to be reassigned. The final assignment of these publications to a research area is the same as the preliminary assignment. In other words, if $x_i^{(l)} \in S^{(l)}$, then $y_i^{(l)} = x_i^{(l)}$. Publications with a preliminary assignment to a research area that is not in the set $S^{(l)}$ are reassigned as follows. For each research area $u$ that is not in the set $S^{(l)}$, we identify the area $v$ in the set $S^{(l)}$ that is most strongly related to area $u$. All publications with a preliminary assignment to area $u$ are then reassigned to area $v$. Hence, if $x_i^{(l)} = u \notin S^{(l)}$, then

$$y_i^{(l)} = \arg\max_{v \in S} \bar{a}_{uv}^{(l)}. \tag{7}$$

---

[3] The same optimization algorithm has also been implemented in the most recent version of our VOSviewer software (Van Eck & Waltman, 2010). However, this software cannot handle very large data sets such as the one used in this paper.

There is one exception. Sometimes a research area $u$ that is not in the set $S^{(l)}$ does not have any relation with areas that are in the set $S^{(l)}$. In that case, publications with a preliminary assignment to area $u$ cannot be reassigned in a proper way. We simply exclude such publications from the classification system.

This completes the description of the second step of our proposed methodology. A summary of this step is provided by the algorithm in Figure 1.

---

**Input:** $c_{ij}, L, r^{(l)}, n_{\min}^{(l)}$ ($i = 1,\ldots,n; j = 1,\ldots,n; l = 1,\ldots,L$)
**Output:** $y_i^{(l)}$ ($i = 1,\ldots,n; l = 1,\ldots,L$)

$[a_{ij}] \leftarrow \text{NormalizeRelatednessScores}([c_{ij}])$
**for** $l \leftarrow L$ **to** $1$ **do**
    Stage 1: $[x_i^{(l)}] \leftarrow \text{MakePreliminaryAssignment}([a_{ij}],[y_i^{(l+1)}],r^{(l)})$
    Stage 2: $[y_i^{(l)}] \leftarrow \text{MakeFinalAssignment}([a_{ij}],[x_i^{(l)}],n_{\min}^{(l)})$
**end for**

---

Figure 1. Algorithm that summarizes the second step of our proposed methodology.

### 2.3. Step 3: Labeling research areas

In the second step of our methodology, the basic structure of a classification system has been built by clustering publications into research areas. In the third step, we finalize the construction of a classification system by assigning labels to the research areas in a system. These labels are obtained by extracting suitable terms from the titles and abstracts of the publications belonging to a research area. A single term is usually not sufficient to clearly indicate what a research area is about. We therefore choose to characterize each research area by a set of terms.

The approach that we take to label the research areas in a classification system consists of the following three stages:
1. *Identification of terms in titles and abstracts of publications*. In this stage, we take the titles and abstracts of all publications included in a classification system and we identify all terms occurring in these titles and abstracts. We first perform part-of-speech tagging (i.e., identification of verbs, nouns, adjectives, etc.). We use the Apache OpenNLP toolkit[4] for this purpose. We then apply a linguistic filter to identify noun phrases. Our filter selects all word sequences that consist exclusively of nouns and adjectives and that end with a noun (e.g., *paper*, *visualization*, *interesting result*, and *text mining*, but not *degrees of freedom* and *highly cited publication*). Finally, we convert plural noun phrases into singular ones. The singular noun phrases serve as our terms.

---

[4] This toolkit is available at http://incubator.apache.org/opennlp/.

2. *Calculation of term relevance scores*. In this stage, we first collect for each research area in a classification system the terms that occur in the titles and abstracts of the publications belonging to the research area. We then calculate a relevance score for each term in a research area. The idea is that terms with a higher relevance score provide a better indication of what a research area is about.

   Suppose we have a term $t$ in research area $u$ at level $l$ in a classification system. Suppose research area $u$ is part of research area $v$ at level $l-1$ in the classification system.[5] We then calculate the relevance score of term $t$ in research area $u$ as $n_{ut}/(n_{vt}+m)$, where $m$ denotes a parameter and where $n_{ut}$ and $n_{vt}$ denote the number of publications in, respectively, area $u$ and area $v$ in which term $t$ occurs. Our calculation of term relevance scores is based on two considerations. On the one hand, the frequency of occurrence of term $t$ in area $u$ relative to the frequency of occurrence of term $t$ in area $v$ can be regarded as an indication of the relevance of term $t$ to area $u$. On the other hand, the absolute frequency of occurrence of term $t$ in area $u$ can also be regarded as an indication of term $t$'s relevance. Our calculation aims to find a balance these two considerations. The parameter $m$ determines how the trade-off between the two considerations is made. In this paper, we use $m = 25$.
3. *Selection of the most relevant terms*. In this final stage, we select the most relevant terms for each research area in a classification system. For instance, for each research area five terms may be selected. In principle, the selection of the most relevant terms is done based on the relevance scores calculated in the previous stage. However, in some research areas, it may be that some of the terms with the highest relevance scores are very similar to each other (e.g., *library* and *librarian* or *peer review* and *peer reviewer*). In that case, from the set of similar terms, we include in our selection only the one with the highest relevance score.[6] The other terms are not included and are replaced by terms with a lower relevance score.

The description of the third and last step of our proposed methodology is now complete. A large-scale application of our methodology is presented in the next two sections.

## 3. Application

We used the methodology introduced in the previous section to construct a classification system based on publications in the Web of Science database in the period 2001–2010. All publications of the document types *article*, *letter*, and *review* in the sciences and the social sciences were included. Publications in the arts and humanities were not included. The total number of publications based on which the classification system was constructed is 10.2 million. There are 97.6 million citation relations between these publications.

Our aim was to construct a classification system that consists of three levels: A first level of 10 to 20 broad disciplines, a second level of 500 to 1000 fields, and a third level of 20,000 to 25,000 small subfields. Table 1 lists the parameter values that we used. We spent a considerable amount of time on selecting these values. Most time was needed to find suitable values for the parameters at the highest level of the classification system (i.e., $r^{(1)}$ and $n_{\min}^{(1)}$). Some values for these parameters led to results that we did not consider satisfactory (e.g., neuroscience and social sciences together in the same research area), and some fine-tuning of the parameter values was needed to avoid such results. Using the parameter values in Table 1, a classification system was obtained that consists of 20 research areas at the first level, 672 research areas at the second level, and 22,412 research areas at the third level. We do not claim that our choice of parameter values is in some sense optimal. The parameter values in Table 1 just serve to illustrate the methodology introduced in this paper. In the end, the choice of parameter values should be guided by the purpose for which a classification system is intended to be used.

---

[5] If $l = 1$, we define research area $v$ as the set of all publications included in the classification system.
[6] To measure the similarity of two terms, we calculate both the average length of the terms and the length of their longest common subsequence. Our measure of similarity is obtained by dividing the latter length by the former one.

Table 1. Parameter values used to construct our classification system.

| | |
|---|---|
| $L$ | 3 |
| $r^{(1)}$ | $8 \times 10^{-8}$ |
| $r^{(2)}$ | $2 \times 10^{-6}$ |
| $r^{(3)}$ | $5 \times 10^{-5}$ |
| $n_{\min}^{(1)}$ | 120,000 |
| $n_{\min}^{(2)}$ | 5,000 |
| $n_{\min}^{(3)}$ | 50 |

Finally, let us make a few comments on the way in which we performed our calculations. To maximize the quality function in (4), our optimization algorithm was run 500 times at the lowest level of the classification system and 10,000 times at the other two levels. The optimization algorithm was written in the C language. All other calculations were programmed in MATLAB. A computer with 64 GB internal memory was used for the calculations. Five runs of the optimization algorithm were performed in parallel on this computer. Overall, it took the computer between four and five days to complete all calculations.

## 4. Results

The classification system that we have constructed consists of three levels, and we therefore split the discussion of the system in three subsections, one for each level of the system. We first discuss the highest level (i.e., level 1) of the classification system, we then discuss the middle level (i.e., level 2), and finally we discuss the lowest level (i.e., level 3). Of the 10.2 million publications with which we started the construction of the classification system, 0.8 million could not be included in the system. We discuss these excluded publications in a separate subsection. At the end of this section, we take a more detailed look at the classification system by focusing specifically on publications from a single journal. The journal that we use is the *Journal of the American Society for Information Science and Technology*.

Throughout this section, we use the following notation to refer to the research areas in our classification system:
- Research area $x$: A research area at level 1 of the system.
- Research area $x.y$: A research area at level 2 of the system. This area is a subarea of area $x$ at level 1.

- Research area *x.y.z*: A research area at level 3 of the system. This area is a subarea of area *x.y* at level 2.

We note that the entire classification system is available online. Both the assignment of publications to research areas and the labeling of the research areas can be downloaded from www.ludowaltman.nl/classification_system/.

**4.1. Level 1**

At level 1, our classification system consists of 20 research areas. Figure 2 shows the distribution of publications over these areas. The average number of publications per research area is about 470,000. The largest research area includes almost 1.34 million publications. The smallest research area covers about 130,000 publications.

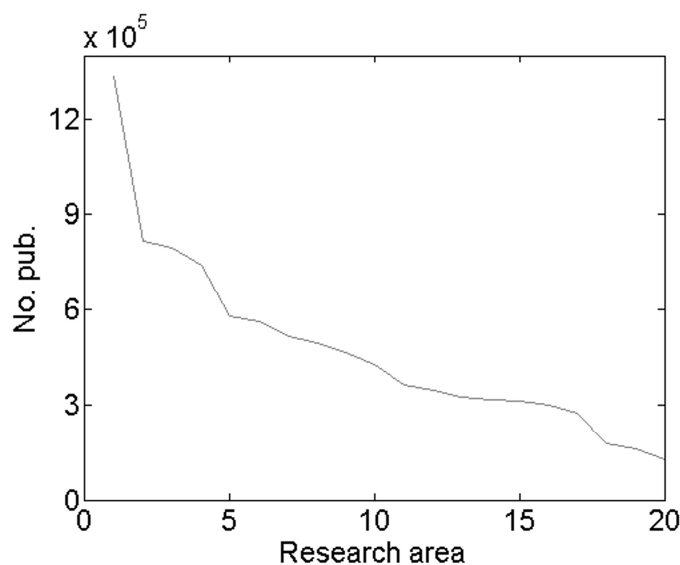

Figure 2. Distribution of publications over the 20 research areas at level 1 of our classification system.

To label the 20 research areas at level 1 of our classification system, we did not use the automated approach discussed in Subsection 2.3. At the highest level of our classification system, we wanted to manually determine suitable labels for our research areas. This turned out to be quite difficult. One might expect to have research areas that correspond closely with well-known broad scientific disciplines such as chemistry, computer science, engineering, mathematics, physics, social sciences, etc. However, we found only a partial correspondence between our research areas and these traditional disciplines. In itself, we believe this to be an interesting result. It may be seen as an indication that traditional disciplines such as those just mentioned only partly reflect the actual organization of today's scientific research. For the purpose of labeling our research areas, however, the imperfect correspondence between our research areas and traditional scientific disciplines created a difficulty. For instance, based on our research areas, we could not make a sufficiently clear distinction between disciplines such as astronomy, chemistry, engineering, materials science, and physics, and we therefore labeled the various areas related to these disciplines simply as *Physical sciences 1*, *Physical sciences 2*, etc. Table 2 lists the labels of all 20 research areas at level 1 of our classification system. As can be seen, many areas have general labels that need further refinement. Refining these labels requires help from experts with a broad overview of the scientific literature in specific disciplines.

Table 2. Labels of the 20 research areas at level 1 of our classification system.

| 1 | Biomedical sciences 1 | 11 | Food and agricultural sciences |
|---|---|---|---|
| 2 | Environmental sciences 1 | 12 | Environmental sciences 2 |
| 3 | Physical sciences 1 | 13 | Physical sciences 3 |
| 4 | Social and health sciences | 14 | Medical sciences 3 |
| 5 | Mathematics and computer science | 15 | Physical sciences 4 |
| 6 | Physical sciences 2 | 16 | Physical sciences 5 |
| 7 | Cognitive sciences | 17 | Physical sciences 6 |
| 8 | Biomedical sciences 2 | 18 | Medical sciences 4 |
| 9 | Medical sciences 1 | 19 | Physical sciences 7 |
| 10 | Medical sciences 2 | 20 | Earth sciences |

To illustrate the difficulty of labeling the research areas, we focus in more detail on the areas labeled *Physical sciences 1* and *Physical sciences 2*. For both areas, Table 3 reports the ten subject categories in Web of Science with which they have most overlap. *Physical sciences 1* has most overlap with physics subject categories, but it also has a considerable overlap with chemistry and materials science categories. This indicates the difficulty of finding a suitable label for this research area. *Physical sciences 2* is clearly dominated by chemistry research, and perhaps it could therefore be relabeled as *Chemistry*. Nevertheless, a considerable number of publications belonging to chemistry subject categories in Web of Science do not belong to *Physical sciences 2* in our classification system. Hence, without help from domain experts, it remains difficult to determine whether relabeling *Physical sciences 2* as *Chemistry* would be correct.

Table 3. Overlap of the research areas labeled *Physical sciences 1* and *Physical sciences 2* in our classification system with subject categories in Web of Science. For each combination of a research area and a subject category, the percentage publications in the research area that belong to the subject category is reported. Publications belonging to multiple subject categories are counted fractionally.

| Physical sciences 1 | | Physical sciences 2 | |
|---|---|---|---|
| Physics, applied | 17.1% | Chemistry, multidisc. | 18.6% |
| Physics, condensed matter | 14.5% | Chemistry, organic | 18.5% |
| Materials science, multidisc. | 10.6% | Chemistry, inorganic & nuclear | 10.3% |
| Chemistry, physical | 9.0% | Chemistry, physical | 9.4% |
| Physics, multidisc. | 6.1% | Crystallography | 6.8% |
| Chemistry, multidisc. | 4.3% | Physics, atomic, molecular & chem. | 5.1% |
| Optics | 3.8% | Biochemistry & molecular biology | 3.1% |
| Engineering, electrical & electronic | 3.7% | Polymer science | 2.6% |
| Electrochemistry | 2.9% | Materials science, multidisc. | 2.0% |
| Materials science, ceramics | 2.1% | Physics, applied | 1.9% |

**4.2. Level 2**

Level 2 of our classification system consists of 672 research areas. The smallest and the largest area include about 5,000 and 48,000 publications, respectively. The average number of publications per area is about 14,000. Figure 3 shows the distribution of publications over the 672 areas.

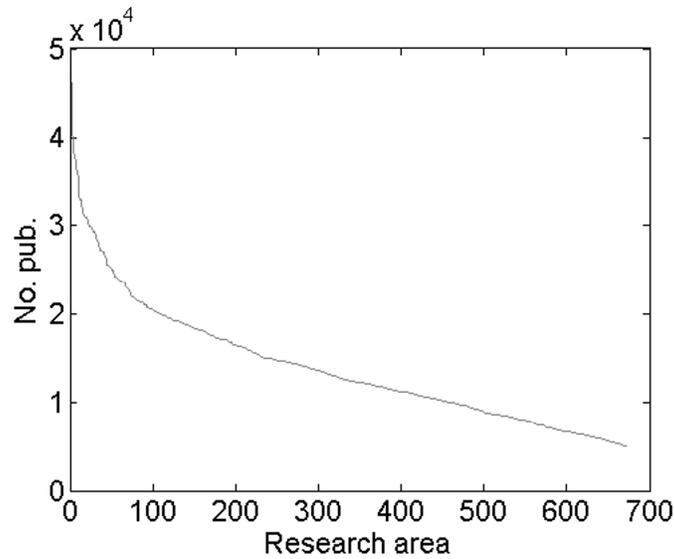

Figure 3. Distribution of publications over the 672 research areas at level 2 of our classification system.

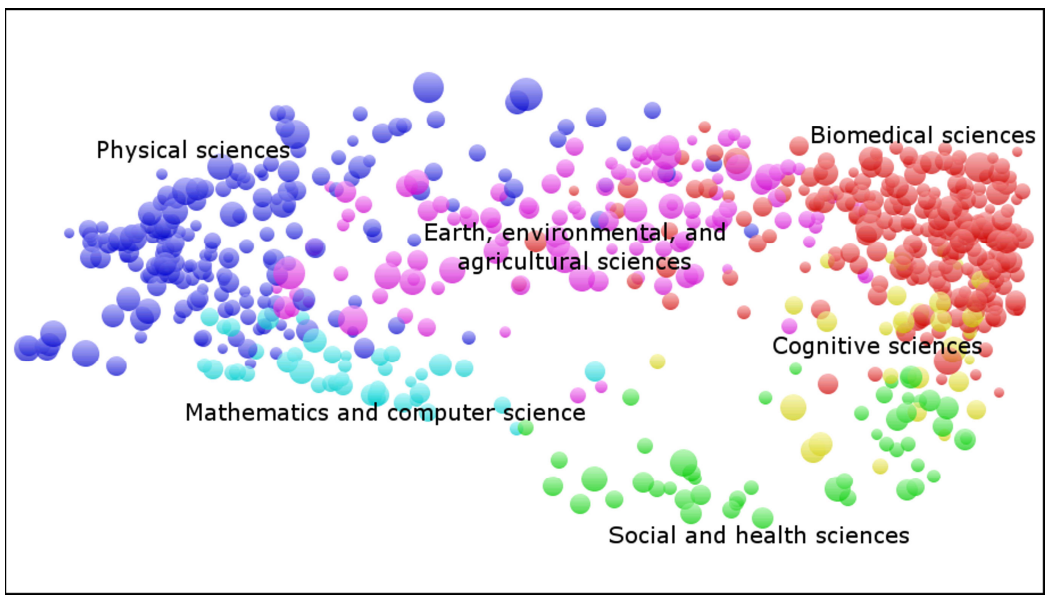

Figure 4. Map of the 672 research areas at level 2 of our classification system. For ease of interpretation, the research areas have been grouped into six categories, which are indicated by colors. Each category corresponds with one or more research areas at level 1 of our classification system.

A visualization of the 672 research areas at level 2 of our classification system is provided in Figure 4. The map in Figure 4 was produced using our VOSviewer software (Van Eck & Waltman, 2010). An interactive version of the map, which offers much more detail, is available at www.ludowaltman.nl/classification_system/. The map was constructed in such a way that strongly related research areas are located close to each other, while research areas that do not have a strong relation are located further away from each other (Van Eck, Waltman, Dekker, & Van den Berg, 2010). The strength of the relation between two research areas was determined based on the number of direct citation relations between publications in the two areas.

The map in Figure 4 has a kind of circular structure. The biomedical sciences are located in the upper right part of the map. Moving counterclockwise from the biomedical sciences, we first observe the earth, environmental, and agricultural sciences in the upper middle part of the map. We then observe the physical sciences in the upper left part of the map. Next, moving downward, we observe mathematics and computer science, and moving further to the lower right part of the map, we observe the social and health sciences. Finally, the circle is closed by the cognitive sciences, which are located in between the social and health sciences and the biomedical sciences. The general structure of science shown in Figure 4 is similar to what has been observed in earlier studies in which for instance journals or Web of Science subject categories were mapped (Klavans & Boyack, 2009; Van Eck & Waltman, 2010).

One application for which our classification system could be used is to detect hot research areas. We define a hot research area as a research area that is quickly expanding in terms of publication output. To detect the hottest research areas at level 2 of our classification system, we calculated for each research area the average publication year of the publications belonging to the area. Table 4 lists the three research areas with the highest average publication year.[7] For each area, the table shows the three most important journals and five characteristic terms. The terms were selected using our labeling approach discussed in Subsection 2.3.

Table 4. The three hottest research areas at level 2 of our classification system.

| Research area | No. pub. | Avg. pub. year | Journals/terms |
|---|---|---|---|
| 3.47 | 6,911 | 2008.3 | *Journals*: Physical Review B; Physical Review Letters; Applied Physics Letters<br>*Terms*: bilayer graphene; graphene oxide; Dirac point; epitaxial graphene; topological insulator |
| 1.81 | 8,290 | 2007.5 | *Journals*: Nucleic Acids Research; PNAS; RNA<br>*Terms*: microRNAs; miRNA expression; miRNA function; mature miRNA; miRNA biogenesis |
| 8.11 | 17,405 | 2006.8 | *Journals*: Vaccine; Journal of Virology; Emerging Infectious Diseases<br>*Terms*: oseltamivir; hMPV; H1N1; RSV infection; H5N1 viruse |

As can be seen in Table 4, the three hottest research areas at level 2 of our classification system are in the fields of physics, molecular biology, and virology. In physics, graphene research constitutes a hot research area. Of the 6,911 publications in this area, 75% appeared in 2008, 2009, or 2010. Notice that in 2010 the Nobel Prize in Physics was awarded to Andre Geim and Konstantin Novoselov "for groundbreaking experiments regarding the two-dimensional material graphene".[8] Geim and Novoselov have, respectively, 70 and 68 publications in our graphene research area, and they are the first two authors of the three most highly cited publications in this area. In molecular biology, microRNA research turns out to be a hot area. The hot research area in virology seems to deal mainly with influenza viruses.

---

[7] The average publication year of all 9.5 million publications included in our classification system is 2006.0.
[8] See www.nobelprize.org/nobel_prizes/physics/laureates/2010/.

**4.3. Level 3**

Level 3 of our classification system consists of 22,412 research areas. On average, each area includes 422 publications. However, as can be seen in Figure 5, there are large differences in the size of the areas. The largest area includes 4,170 publications, while the smallest area covers only 50 publications.

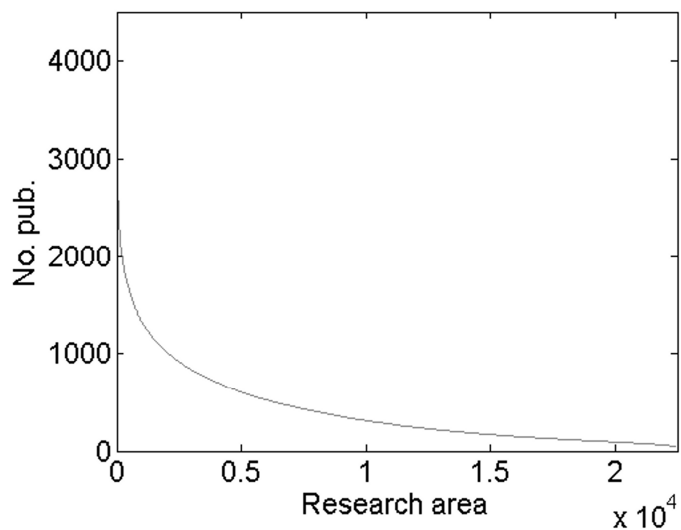

Figure 5. Distribution of publications over the 22,412 research areas at level 3 of our classification system.

Table 5. The three hottest research areas at level 3 of our classification system.

| Research area | No. pub. | Avg. pub. year | Journals/terms |
|---|---|---|---|
| 3.16.2 | 1,999 | 2009.2 | *Journals*: Physical Review B; Physica C; Physical Review Letters<br>*Terms*: iron; iron pnictide; BaFe2As2; Fe1 xCox; iron pnictide superconductor |
| 3.47.2 | 1,199 | 2009.2 | *Journals*: Carbon; ACS Nano; Journal of Physical Chemistry C<br>*Terms*: graphene oxide; composite; water; chemical reduction; preparation |
| 6.12.1 | 2,326 | 2009.0 | *Journals*: Acta Crystallographica E; Zeitschrift für Kristallographie - New Crystal Structures; Acta Crystallographica C<br>*Terms*: rms; Sn IV atom; deviation; inversion dimer; Ru atom |

Table 5 lists the three hottest research areas at level 3 of our classification system. Like we did above for level 2 of our system, we detected these areas based on the average publication year of their publications. All three areas are in the physical sciences, and they all have more than 90% of their publications in the period 2008–2010. Research area 3.16.2, which has even more than 98% of its publications in this period, deals with high-temperature superconductivity. Research area 3.47.2 deals with graphene research. This area is a subarea of the graphene area discussed in the previous subsection. The third research area, area 6.12.1, deals with a topic in the field of crystallography. This area includes the most highly cited publication in our classification system.[9] Despite its relatively recent publication date (January 2008), this publication has already been cited more than 20,000 times. The publication is about a set of computer programs used in crystallography research.

To get some more insight into the characteristics of the research areas at level 3 of our classification system, we consider a single research area in more detail. Figure 6 shows a map of the 417 publications belonging to research area 4.30.10. The 1,197 citation relations between these publications are displayed as well. The map was produced using our VOSviewer software. An interactive version of the map can be found at www.ludowaltman.nl/classification_system/. As can be seen in Figure 6, research area 4.30.10 deals with the topic of mapping or visualization of science. This is also indicated by the terms that have been selected to label the area. These terms are *ACA* (abbreviation of author co-citation analysis), *research front*, *similarity measure*, *information visualization*, and *map*.

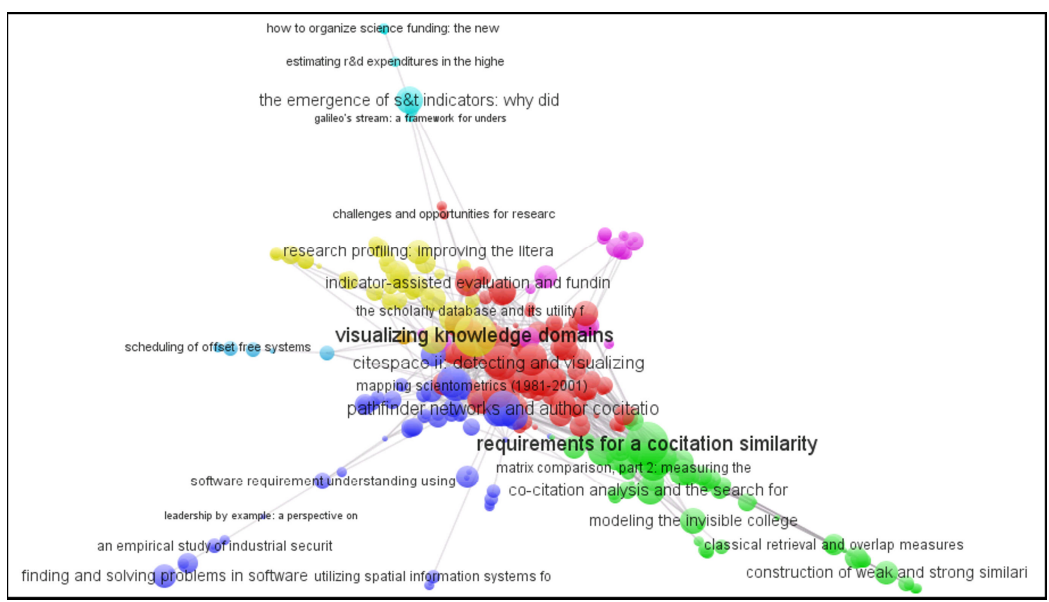

Figure 6. Map of the 417 publications belonging to research area 4.30.10 of our classification system.

The map in Figure 6 shows that research area 4.30.10 has a core of strongly related publications and a periphery of publications that are only weakly related to other publications in the area. This is a quite typical structure for many other research areas as well. Looking in more detail at the map (which can best be done using the interactive version of the map), it can be seen that the publications in the core are almost all in some way related to the topic of mapping or visualization of science. However, some publications in the periphery of the map have no clear relation with this topic. These publications for instance deal with the quality of computer software (in the lower left part of the map) or with science policy related topics (in the upper part of the map). The assignment of these publications to research area 4.30.10 is perhaps understandable from the point of view of the citation relations between publications, but from a substantive point of view it is not satisfactory. The unsatisfactory way in which these publications have been classified seems to be a consequence of the fact that our methodology uses only direct citation relations to determine the relatedness of publications. Some of the unsatisfactory assignments of publications to research areas can probably be avoided by also using indirect citation relations (e.g., bibliographic coupling relations) or relations based on shared words in titles and abstracts. We will come back to this issue at the end of this paper.

---

[9] This the following publication: Sheldrick, G.M. (2008). A short history of SHELX. *Acta Crystallographica Section A*, *64*(1), 112–122.

**4.4. Excluded publications**

As already mentioned, of the 10.2 million publications with which we started the construction of our classification system, 0.8 million could not be included in the system. We now look at these excluded publications.

In the methodology introduced in this paper, a publication can be included in a classification system only if in the citation network discussed in Subsection 2.1 the publication belongs to a connected component that consists of at least $n_{\min}^{(L)}$ publications. In other words, a publication must have direct or indirect citation relations with at least $n_{\min}^{(L)} - 1$ other publications. In our case, this means that a publication must be related, either directly or indirectly, to at least 49 other publications. The 0.8 million excluded publications do not meet this criterion. In fact, 91% of these publications turn out to have no relations at all with other publications. We note that the publications included in our classification system all belong to a single very large connected component of the citation network.

Table 6. Percentage excluded publications per publication year (left column) and the ten Web of Science subject categories with the highest percentage excluded publications (right column).

| Perc. excluded pub. per year | | Perc. excluded pub. per subject category | |
|---|---|---|---|
| 2001 | 14.3% | Engineering, marine | 51.8% |
| 2002 | 11.5% | Political science | 34.7% |
| 2003 | 9.5% | Engineering, aerospace | 33.7% |
| 2004 | 7.9% | Cultural studies | 31.9% |
| 2005 | 7.0% | Business, finance | 29.4% |
| 2006 | 6.1% | Area studies | 28.7% |
| 2007 | 5.9% | Engineering, petroleum | 27.9% |
| 2008 | 5.8% | Materials science, paper & wood | 26.3% |
| 2009 | 5.5% | Social issues | 22.0% |
| 2010 | 5.2% | Information science & library science | 21.8% |

Table 6 reports the percentage excluded publications per publication year. The table also lists the ten Web of Science subject categories with the highest percentage excluded publications. Overall, 7.6% of all publications were excluded from our classification system. As can be seen in the table, excluded publications are overrepresented in earlier publication years. This is because many references in publications published in earlier years go back before 2001 and therefore are not taken into account in our methodology. Not surprisingly, excluded publications are also overrepresented in fields with a low citation density, in particular in engineering research and in the social sciences. Many excluded publications in these fields seem to have appeared in special types of journals, such as national scientific journals, trade journals, and popular magazines. It further turns out that 16.2% of the 0.8 million excluded publications are of the Web of Science document type *letter*, while overall just 3.8% of the publications are of this document type.

We emphasize that there are various possibilities to make sure that in the end all or almost all publications are included in a classification system. For instance, publications currently excluded from our classification system could be added to the system based on an analysis of the words occurring in their titles and abstracts.

**4.5. Classification of *JASIST* publications**

We now take a more detailed look at specific areas of our classification system. To do so, we focus on publications from the *Journal of the American Society for Information Science and Technology* (*JASIST*). *JASIST* is a leading journal in the field of information science. We choose to focus on publications from *JASIST* because of our familiarity with this journal. Moreover, many readers of this paper will probably be familiar with *JASIST* as well.

In the period 2001–2010, *JASIST* published 1,499 publications classified as *article*, *letter*, or *review* in Web of Science. Of these publications, 62 were excluded from our classification system. Table 7 reports the distribution of the remaining 1,437 publications over the research areas at level 1 of our classification system. Almost 97% of the *JASIST* publications turn out to belong to either the area labeled *Social and health sciences* or the area labeled *Mathematics and computer science*. This is in line with what one may expect, since publications in *JASIST* can all or almost all be classified as either social science research or computer science research, or as a combination of these two types of research.

Table 7. Distribution of 1,437 *JASIST* publications over the research areas at level 1 of our classification system.

| Social and health sciences | 997 |
|---|---|
| Mathematics and computer science | 394 |
| Environmental sciences 1 | 8 |
| Physical sciences 3 | 8 |
| Cognitive sciences | 7 |
| Biomedical sciences 1 | 6 |
| Biomedical sciences 2 | 4 |
| Other research areas | 13 |

What is perhaps more surprising in Table 7 is that there are 46 *JASIST* publications which do not belong to either *Social and health sciences* or *Mathematics and computer science*. We looked at a number of these publications in more detail. It turns out that some publications have been clearly misclassified. Like the misclassifications discussed at the end of Subsection 4.3, these misclassifications seem to be a consequence of the fact that publications sometimes have only a very small number of direct citation relations with other publications. However, we also found that some publications have been correctly assigned to research areas different from *Social and health sciences* and *Mathematics and computer science*. For instance, Table 8 lists the three most frequently cited *JASIST* publications in the area labeled *Physical sciences 3*. At level 2 of our classification system, these three publications all belong to area 13.5. This area covers the field of network analysis, which is a field that receives a lot of attention in physics journals. Looking in more detail at the publications in Table 8, we believe that their assignment to the field of network analysis is quite sensible. Hence, in specific cases, assigning *JASIST* publications to research areas different from *Social and health sciences* and *Mathematics and computer science* seems perfectly reasonable.

Table 8. The three most frequently cited *JASIST* publications in the research area labeled *Physical sciences 3* at level 1 of our classification system.

- Matia et al. (2005). Scaling phenomena in the growth dynamics of scientific output.
- Panzarasa et al. (2009). Patterns and dynamics of users' behavior and interaction: Network analysis of an online community.
- Havemann et al. (2005). Firm-like behavior of journals? Scaling properties of their output and impact growth dynamics.

Table A1 in the appendix lists the five research areas at level 2 of our classification system with the largest number of *JASIST* publications. For each area, the table shows the three *JASIST* publications that have received most citations. Three of the five areas are subareas of the *Social and health sciences* area at level 1 of our classification system. The other two areas are subareas of the *Mathematics and computer science* area. Research area 4.30 includes more than half of the publications of *JASIST*. This area covers large parts of the field of information science. The other two subareas of the *Social and health sciences* area include a smaller number of *JASIST* publications. These two areas cover specific information science topics not covered by area 4.30. The two subareas of the *Mathematics and computer science* area cover information retrieval related topics.[10]

We now turn to level 3 of our classification system. Table A2 in the appendix lists the five research areas at this level with the largest number of *JASIST* publications. These areas all turn out to be subareas of area 4.30 at level 2 of our classification system. A research area at level 3 usually covers a single well-defined topic. Table A2 suggests that the five most important topics addressed by *JASIST* in the period 2001–2010 may be labeled as follows:

- Searching behavior, in particular on the Web (area 4.30.2).
- Bibliometric indicators of scientific performance (area 4.30.1).
- Mapping or visualization of science (area 4.30.10).
- Webometrics (area 4.30.6).
- Foundations of information science (area 4.30.4).

## 5. Conclusion and future research

In this paper, we have introduced a new methodology for constructing a classification system of science. Most classification systems are defined at the level of journals, but our proposed methodology works at the level of individual publications. In the application that we have presented, a classification system was produced that includes almost ten million publications. This exceeds other recently published studies (e.g., Boyack & Klavans, 2010; Boyack et al., 2011; Klavans & Boyack, 2010) and makes it possible to cover essentially all Web of Science indexed publications in a ten-year time period.

---

[10] Table A1 also shows for each research area the terms that have been selected to label the area. As can be seen in the table, these terms do not always properly reflect what an area is about. This is especially clear in the case of area 4.2. Our impression is that the labeling approach discussed in Subsection 2.3 yields more satisfactory results at level 3 of our classification system than at level 2. In general, at higher levels of aggregation, it seems more difficult to automatically identify suitable labels for a research area.

A noteworthy feature of our methodology is its transparency and relative simplicity. Our methodology is fully documented in this paper and consists of a limited number of steps that are all fairly easy to understand. There are only a small number of parameters for which values need to be chosen manually. Each level of a classification system results in just two additional parameters: The resolution parameter and the minimum number of publications per research area. The clustering software that we use in our methodology is freely available online. Anyone can use this software for his own purposes. The requirements of our methodology in terms of computing time and memory usage are relatively modest, although large-scale applications such as the one presented in this paper may be too demanding for a standard desktop computer.

To determine the relatedness of publications, our methodology relies exclusively on direct citation relations. This is the main reason for the modest computing and memory requirements of our methodology. At the same time, however, this can also be considered the main weakness of our approach. As we have seen in Section 4, a substantial proportion of all publications do not have sufficient direct citation relations to be included in a classification system. Also, some publications with only a few direct citation relations can be included in a classification system, but they may be assigned to an incorrect research area. In principle, it should be possible to increase the coverage and the accuracy of our methodology by using a more sophisticated measure of the relatedness of publications. In addition to direct citation relations, such a measure could also take into account indirect citation relations. Especially the use of bibliographic coupling relations may lead to significant improvements. An even more sophisticated approach could be to use a measure that combines citation relations with relations based on shared words in titles and abstracts (e.g., Ahlgren & Colliander, 2009; Boyack & Klavans, 2010; Janssens, Glänzel, & De Moor, 2008). However, there is a crucial practical limitation. The use of a more sophisticated measure of the relatedness of publications increases the computing and memory requirements of our methodology. In large-scale applications such as the one presented in this paper, many sophisticated measures of relatedness therefore cannot be used in a straightforward way. For instance, within a set of ten million publications, there may be billions of bibliographic coupling relations, and taking into account all these relations is likely to be too demanding, both in terms of computing time and in terms of memory usage. An attractive alternative approach may be to use a more sophisticated measure of relatedness, but to take into account only the strongest relations between publications. This is somewhat similar to the 'top *n* similarity approach' introduced by Klavans and Boyack (2006).

In addition to the use of more sophisticated measures of the relatedness of publications, there are a number of other issues that need to be addressed in future research:

- *Improve the labeling of research areas*. There seems to be room for improving the approach that we take to label the research areas in a classification system. Improvement is needed especially at higher levels of aggregation. One possibility may be to label research areas using journal title words.
- *Allow for overlap of research areas*. Our methodology assigns each publication to a single research area. Because of this, there is no overlap of research areas. For some purposes, it may be desirable to allow research areas to overlap each other. In that case, publications related to multiple fields or multiple topics could be assigned to more than one research area.

- *Evaluate the accuracy of the proposed methodology in a more rigorous way*. In this paper, we did not provide a rigorous evaluation of the accuracy of our methodology. Performing such an evaluation is difficult because of the lack of a 'golden standard'. In future work, a more rigorous evaluation of the accuracy of our methodology may be performed based on expert feedback.
- *Use the proposed methodology to construct a journal-level classification system*. Our focus in this paper has been on constructing a publication-level classification system. Nevertheless, our methodology may also be useful for constructing a classification system at the level of journals. Directly applying our methodology to journals instead of publications may be problematic because of the multidisciplinary nature of some journals. An alternative approach could be to first construct a classification system at the level of publications and to then derive a journal-level system from this publication-level system.

## Acknowledgment

We would like to thank Ed Noyons for his feedback on the classification system that we have produced.

# Appendix

Table A1. The five research areas at level 2 of our classification system with the largest number of *JASIST* publications.

| | | |
|---|---|---|
| *Area*: 4.30 | *Terms*: h index; academic library; document supply; electronic journal; digital library | *No. JASIST pub.*: 758 |

*Most frequently cited JASIST publications*:
- Spink et al. (2001). Searching the Web: The public and their queries.
- Meho & Yang (2007). Impact of data sources on citation counts and rankings of LIS faculty: Web of science versus Scopus and Google Scholar.
- Jansen & Pooch (2001). A review of Web searching studies and a framework for future research.

| | | |
|---|---|---|
| *Area*: 5.20 | *Terms*: biomedical literature; CLEF; UMLS; MEDLINE abstract; recommender system | *No. JASIST pub.*: 221 |

*Most frequently cited JASIST publications*:
- Srinivasan (2004). Text mining: Generating hypotheses from MEDLINE.
- Weeber et al. (2001). Using concepts in literature-based discovery: Simulating Swanson's Raynaud-fish oil and migraine-magnesium discoveries.
- Liben-Nowell & Kleinberg (2007). The link-prediction problem for social networks.

| | | |
|---|---|---|
| *Area*: 4.2 | *Terms*: enterprise resource planning; ERP system; ERP implementation; TQM; new product development | *No. JASIST pub.*: 74 |

*Most frequently cited JASIST publications*:
- Alavi & Tiwana (2002). Knowledge integration in virtual teams: The potential role of KMS.
- McInerney (2002). Knowledge management and the dynamic nature of knowledge.
- Kostoff et al. (2001). Citation mining: Integrating text mining and bibliometrics for research user profiling.

| | | |
|---|---|---|
| *Area*: 5.3 | *Terms*: face recognition; minutiae; query image; image retrieval; partial occlusion | *No. JASIST pub.*: 66 |

*Most frequently cited JASIST publications*:
- Choi & Rasmussen (2003). Searching for images: The analysis of users' queries for image retrieval in American history.
- Chen (2001). An analysis of image queries in the field of art history.
- Jorgensen & Jorgensen (2005). Image querying by image professionals.

| | | |
|---|---|---|
| *Area*: 4.10 | *Terms*: internet addiction; service failure; violent video game; technology acceptance model; New York University | *No. JASIST pub.*: 65 |

*Most frequently cited JASIST publications*:
- Rieh (2002). Judgment of information quality and cognitive authority in the Web.
- Wathen & Burkell (2002). Believe it or not: Factors influencing credibility on the Web.
- Thelwall (2008). Social networks, gender, and friending: An analysis of MySpace member profiles.

Table A2. The five research areas at level 3 of our classification system with the largest number of *JASIST* publications.

| | | |
|---|---|---|
| *Area*: 4.30.2 | *Terms*: query; web searching; searcher; information behaviour; search task | *No. JASIST pub.*: 246 |

*Most frequently cited JASIST publications*:
- Spink et al. (2001). Searching the Web: The public and their queries.
- Jansen & Pooch (2001). A review of Web searching studies and a framework for future research.
- Borlund (2003). The concept of relevance in IR.

| | | |
|---|---|---|
| *Area*: 4.30.1 | *Terms*: h index; journal impact factor; Hirsch; self citation; Scopus | *No. JASIST pub.*: 230 |

*Most frequently cited JASIST publications*:
- Meho & Yang (2007). Impact of data sources on citation counts and rankings of LIS faculty: Web of science versus Scopus and Google Scholar.
- Cronin & Meho (2006). Using the h-index to rank influential information scientists.
- Bornmann & Daniel (2007). What do we know about the h index?

| | | |
|---|---|---|
| *Area*: 4.30.10 | *Terms*: ACA; research front; similarity measure; information visualization; map | *No. JASIST pub.*: 63 |

*Most frequently cited JASIST publications*:
- Ahlgren et al. (2003). Requirements for a cocitation similarity measure, with special reference to Pearson's correlation coefficient.
- Chen (2006). CiteSpace II: Detecting and visualizing emerging trends and transient patterns in scientific literature.
- White (2003). Pathfinder networks and author cocitation analysis: A remapping of paradigmatic information scientists

| | | |
|---|---|---|
| *Area*: 4.30.6 | *Terms*: link analysis; hyperlink; web link; inlink; URLs | *No. JASIST pub.*: 61 |

*Most frequently cited JASIST publications*:
- Thelwall (2001). Extracting macroscopic information from Web links.
- Thelwall (2002). Conceptualizing documentation on the Web: An evaluation of different heuristic-based models for counting links between university Web sites.
- Koehler (2002). Web page change and persistence: A four-year longitudinal study.

| | | |
|---|---|---|
| *Area*: 4.30.4 | *Terms*: knowledge organization; epistemology; classification scheme; GIS; Hjorland | *No. JASIST pub.*: 52 |

*Most frequently cited JASIST publications*:
- Hjorland (2002). Epistemology and the socio-cognitive perspective in information science.
- Bates (2006). Fundamental forms of information.
- Hjorland (2001). Towards a theory of aboutness, subject, topicality, theme, domain, field, content ... and relevance.